\documentstyle[prl,aps,multicol]{revtex}

\begin{document}
\draft
\title{Density functional formalism in the canonical ensemble}
\author{J.A. Hernando\cite{jah}}
\address{Dept. of Phys., Comisi\'on Nacional de Energ\'\i a
At\'omica,\\
Av. del Libertador 8250, 1429 Buenos Aires, Argentina}
\author{L. Blum\cite{lb}}
\address{Dept. of Phys., POB 23343, Univ. of Puerto Rico,\\
Rio Piedras, PR 00931-3343, USA}
\date{\today }
\maketitle

\begin{abstract}
Density functional theory, when applied to systems with $T\neq 0$, is based
on the grand canonical  extension of the Hohenberg-Kohn-Sham theorem due to
Mermin (HKSM theorem). While a straightforward canonical ensemble
generalization fails,  work in nanopore systems
could certainly benefit from such extension.
We show that, if the asymptotic behaviour of the
canonical distribution functions is taken into account, the HKSM\ theorem
can be extended to the canonical ensemble. We generate $N$-modified
correlation and distribution functions hierarchies and prove that, if they
are employed, either a modified external field or the density profiles can
be indistinctly used as independent variables. We also write down the $N$%
-modified free energy functional and prove that its minimum is reached when
the equilibrium values of the new hierarchy are used. This completes the
extension of the HKSM\ theorem.
\end{abstract}


\begin{multicols}{2}

Density functional theory (DFT) is, undoubtedly, one of the more reliable
and established tools in condensed matter theory. It has successfully been
used in a great variety of classical systems \cite{revcl} as, e.g., uniform
and non uniform systems in simple \cite{ros,kros} and general \cite{hupi}
fluids, confined fluids \cite{bryk,gevans}, melting and freezing \cite{mf},
interfaces \cite{if}, etc. as well as in the calculation of electronic
properties in all sort of systems \cite{revel}. This impressive amount of
work directly descends from the pioneering work of Hohenberg, Kohn and Sham 
\cite{hokosh} and its  extension to non-zero temperature for systems
described in the grand canonical ensemble (GCE) by Mermin \cite{merm}.
Roughly speaking, the Hohenberg-Kohn-Sham-Mermin (HKSM) theorem states that,
either the external potential or the density profile can indistinctly be
used as independent variables and that the thermodynamic grand potential
reaches its minimum value when the equilibrium density profile is used.
Therefore, it is in the foundation of all sort of variational principles.
The failure to implement a straightforward canonical ensemble (CE) extension
is already well known, referred to in \cite{gevans},
 and can be traced back to the fixed N constraint (see
eq. (\ref{canorm})). As the GCE is an extension of the CE, an obvious
question is where the need for such an implementation exists. In relation to
that question, we can mention that experiments done in porous glasses (with a mean
pore radius $r\approx 20-30\AA $ )\cite{expor}, simulations \cite{mcpor} and
DFT studies \cite{bryk,gevans} show the interest of having a mesoscopic DFT in
the CE in order to study the statistical mechanics of finite closed systems
like fluids in spherical pores \cite{gevans}.  In this letter we consider a general
mixture of $p$ components (labeled by greek indices) with $N_\alpha $
particles in a box of volume $V$ and under the influence 
of one body, species dependent, external
potentials $V_\alpha ^{(1)}({\bf x})$. Following Ramshaw's work \cite{rams}
we will show how, by stripping  the canonical correlation functions off of
its asymptotic behaviour, an N-modified set of distribution and correlation
function hierarchies (which we also show can be defined by functional
derivatives of the canonical partition function $Q_N$ with respect to a
modified external potential) is introduced and it paves the way to prove the
HKSM theorem in the CE. More specifically, we do that by introducing an $N$%
-modified free energy which is minimized by the $N$-modified density
profiles. Here we present the essential results 
with the pair correlation function case as a prototype and a complete
derivation will be published in a paper in preparation.

As usual \cite{revliq}, the distribution functions and full distribution
functions, the truncated correlation functions associated to them and direct
correlation functions hierarchies ($n_{\mbox{\boldmath $\alpha$}}^{(s)}(\{%
{\bf x}\})$, $\hat n_{\mbox{\boldmath $\alpha$}}^{(s)}(\{{\bf x}\})$, $t_{%
\mbox{\boldmath $\alpha$}}^{(s)}(\{{\bf x}\})$, $\hat t_{%
\mbox{\boldmath
$\alpha$}}^{(s)}(\{{\bf x}\})$, ($c_{\mbox{\boldmath $\alpha$}}^{(s)}(\{{\bf %
x}\})$ respectively) can be defined by functional derivatives as

\end{multicols}

\begin{eqnarray}
n_{\mbox{\boldmath $\alpha$}}^{(s)}(\{{\bf x}\}) &=&\frac{%
\prod_{k=1}^se^{\Phi _{\alpha _k}({\bf x}_k)}}{Q_{{\bf N}}}\frac{\delta ^sQ_{%
{\bf N}}}{\prod_{k=1}^s\delta e^{\Phi _{\alpha _k}({\bf x}_k)}}
\label{dis-us} \\
\hat n_{\mbox{\boldmath $\alpha$}}^{(s)}(\{{\bf x}\}) &=&\frac 1{Q_{{\bf N}}}%
\frac{\delta ^sQ_{{\bf N}}}{\prod_{k=1}^s\delta \Phi _{\alpha _k}({\bf x}_k)}
\label{dismod-us} \\
t_{\mbox{\boldmath $\alpha$}}^{(s)}(\{{\bf x}\}) &=&\prod_{k=1}^se^{\Phi
_{\alpha _k}({\bf x}_k)}\frac{\delta ^s\ln Q_{{\bf N}}}{\prod_{k=1}^s\delta
e^{\Phi _{\alpha _k}({\bf x}_k)}}  \label{corr-us} \\
\hat t_{\mbox{\boldmath $\alpha$}}^{(s)}(\{{\bf x}\}) &=&\frac{\delta ^s\ln
Q_{{\bf N}}}{\prod_{k=1}^s\delta \Phi _{\alpha _k}({\bf x}_k)}=\frac{\delta 
\hat t_{\alpha _1...\alpha _{s-1}}^{(s-1)}(\{{\bf x}_1{\bf ,...,x}_s\})}{%
\delta \Phi _{\alpha s}({\bf x}_s)}  \label{corrmod-us} \\
c_{\mbox{\boldmath $\alpha$}}^{(s)}(\{{\bf x}\}) &=&\beta \frac{\delta
^sF^{exc}}{\prod_{k=1}^s\delta n_{\alpha _k}^{(1)}({\bf x}_k)}=\frac{\delta
c_{\alpha _1...\alpha _{s-1}}^{(s-1)}(\{{\bf x}_1{\bf ,...,x}_s\})}{\delta
n_{\alpha s}^{(1)}({\bf x}_s)}  \label{dcorr-us}
\end{eqnarray}

\begin{multicols}{2}

\noindent where $\mbox{\boldmath $\alpha$}=(\alpha _1,\ldots ,\alpha _s)$, $%
\{{\bf x}\}=({\bf x}_1,\ldots ,{\bf x}_s)$ are the species and coordinate
sets, $\beta =(kT)^{-1}$, $F^{exc}$ is the excess free energy and $\Phi
_\alpha ({\bf x})=-\beta V_\alpha ^{(1)}({\bf x})$. The GCE definition is
formally the same with the grand partition function $\Xi $ replacing $Q_N$.
We will consider that the full distribution and correlation functions $\hat n%
^{(s)},\hat t^{(s)}$ are the fundamental quantities (together with $c^{(s)}$%
) in the sense that they not only pinpoint the problems in the CE, is
through these functiones that we solve them and no new information can be
gotten from the other functions. It is obvious that $n_\alpha ^{(1)}=\hat n%
_\alpha ^{(1)}=t_\alpha ^{(1)}=\hat t_\alpha ^{(1)}$ and that

\begin{equation}
\delta n_\alpha ^{(1)}({\bf x})=\sum_\lambda \int \hat t_{\alpha \lambda
}^{(2)}({\bf x,y})\delta \Phi _\lambda ({\bf y})d{\bf y}  \label{deldis-us}
\end{equation}

\noindent The link with the more usual notation is $\hat t_{\alpha \lambda
}^{(2)}({\bf x,y})=t_{\alpha \lambda }^{(2)}({\bf x,y})+n_\alpha ^{(1)}({\bf %
x})\delta _{\alpha \lambda }\delta ({\bf x-y)}$ where $t_{\alpha \lambda
}^{(2)}({\bf x,y})=n_\alpha ^{(1)}({\bf x})n_\lambda ^{(1)}({\bf y}%
)h_{\alpha \lambda }^{(2)}({\bf x,y})$ and, in this case, the independent
variables are the external fields. If the density profiles $\{n_\alpha
^{(1)}\}$ can also be used as independent variables, then an inverse kernel $%
\hat t_{\eta \alpha }^{(2)-1}({\bf z,x})=\delta \Phi _\eta ({\bf z})/\delta
n_\alpha ^{(1)}({\bf x})$ must exist such that

\begin{equation}
\sum_\alpha \int \hat t_{\eta \alpha }^{(2)-1}({\bf z,x})\hat t_{\alpha
\lambda }^{(2)}({\bf x,y})d{\bf x}=\delta _{\eta \lambda }\delta ({\bf z-y)}
\label{ozfunct}
\end{equation}

\noindent Using the definition of the pair distribution function in both
ensembles \cite{revliq} , it can be obtained that

\end{multicols}

\begin{equation}
\int \hat t_{\alpha \lambda }^{(2)}({\bf x,y})d{\bf y=}\left\{ 
\begin{tabular}{ll}
0 & CE \\ 
$n_\alpha ^{(1)}({\bf x})\delta _{\alpha \lambda }+n_\alpha ^{(1)}({\bf x})%
\frac{\partial \ln \left[ \Xi n_\alpha ^{(1)}/z_\alpha \right] }{\partial
\ln z_\lambda }-\frac{\partial \ln \Xi }{\partial \ln z_\alpha }\frac{%
\partial \ln \Xi }{\partial \ln z_\lambda }$ & GCE
\end{tabular}
\right.   \label{canorm}
\end{equation}

\begin{multicols}{2}

\noindent Eqs. (\ref{ozfunct}) and (\ref{canorm}) are obviously incompatible
in the CE while there is no incompatibility in the GCE; because of the
fluctuations, the nucleus $\hat t_{\alpha \lambda }^{(2)}$ is invertible.
Therefore, the Ornstein-Zernike (OZ)\ equation (obtained when $c_{\eta
\alpha }^{(2)}({\bf z,x})$ is defined by $\hat t_{\eta \alpha }^{(2)-1}({\bf %
z,x})=-c_{\eta \alpha }^{(2)}({\bf z,x})+\delta _{\eta \alpha }\delta ({\bf %
z-x)/}n_\alpha ^{(1)}({\bf x})$) is, mathematically speaking, undefined in
the CE and, within that framework, no rigorous DFT is possible. In other
words, any variational principle formulated in the CE with the canonical
functions hierarchies and having the density profiles as independent
variables does not have a mathematically sound foundation. Specifically,
this is the problem alluded to in \cite{gevans}.

The way out we present in this letter is through stripping the correlation
functions $\hat t^{(s)}$ off of its asymptotic behaviour. This stripping
defines the new, $N$-modified distribution and correlation functions,  $%
\tilde n^{(s)},\tilde t^{(s)}$ respectively. We then show that $\tilde t%
^{(s)}$ is invertible and, from $\tilde t^{(s)-1}$ we define $N$-modified
direct correlation functions $\tilde c^{(s)}$, write down OZ equations that
link $\tilde t^{(s)}$ and $\tilde c^{(s)}$ and, by proving that they can be
also obtained by functional diferentiation of the CE partition function with
respect to modified external fields $\tilde \Phi _\lambda $, we conclude
that $\tilde n^{(s)},\tilde t^{(s)},\tilde c^{(s)}$ constitute a new
hierarchical set of distribution and correlation functions in the CE without
the mathematical restrictions that the more conventional ones have. Lastly,
we  write down a $N$-modified free energy in terms of these $N$-modified
functions and prove that it reaches its minimum value when the equilibrium $N
$-modified density profiles $\{\tilde n_\alpha ^{(1)}\}$ are used. This
completes the proof of the extension of the HKSM\ theorem to the CE. Here we
will sketch the main steps using the pair correlations case as a prototype and
write down the general results. The complete derivation will be reported in
a paper in preparation.

The irreducible two bodies behaviour is exclusively contained in $h_{\alpha
\lambda }^{(2)}$ and we will analyze its asymptotic behaviour. The
conditional probability of finding an $\alpha $ particle in ${\bf x}$ when a 
$\lambda $ particle is fixed in ${\bf y}$ can be written, when ${\bf x}$ and 
${\bf y}$ are very far away, as

\begin{equation}
n_{\alpha \lambda ;\infty }({\bf x}|{\bf y})=n_\alpha ^{(1)}({\bf x})+\frac{%
\partial n_\alpha ^{(1)}({\bf x})}{\partial \rho _\alpha }\Delta \phi
_\lambda ({\bf y})  \label{cond-n2}
\end{equation}

\noindent Here $\Delta \phi _\lambda ({\bf y})$ is an unknown
proportionality factor (obviously $\{\lambda ,{\bf y}\}$ dependent); it is
not the difference in density due to anchoring a $\lambda $ particle. Notice
also that the derivative is with respect to $\rho _\alpha $, not $\rho
_\lambda $. The reasons are not only the particle exchange symmetry, they
are also related to the fact that one of our goals is to obtain a full
hierarchy of $N$-modified functions (otherwise, in eq. (\ref{modfield}) a
modified external field could not be defined). Therefore, the asymptotic
behaviour of $h_{\alpha \lambda }^{(2)}$ can be written in a fully
symmetrical way

\begin{equation}
h_{\alpha \lambda ;\infty }^{(2)}({\bf x,y})=\Gamma _{\alpha \lambda }\frac{%
\partial \ln n_\alpha ^{(1)}({\bf x})}{\partial \rho _\alpha }\frac{\partial
\ln n_\lambda ^{(1)}({\bf y})}{\partial \rho _\lambda }  \label{hasymp}
\end{equation}

\noindent $\Gamma _{\alpha \lambda }$ is an as yet unknown constant. Next,
we define an $N$-modified correlation function $\tilde h_{\alpha \lambda
}^{(2)}$ as the correlation function stripped off of its asymptotic
behaviour. More precisely,

\begin{equation}
h_{\alpha \lambda }^{(2)}({\bf x,y})=\left\{ 
\begin{tabular}{ll}
$\tilde h_{\alpha \lambda }^{(2)}({\bf x,y})+\Gamma _{\alpha \lambda }\frac{%
\partial \ln n_\alpha ^{(1)}({\bf x})}{\partial \rho _\alpha }\frac{\partial
\ln n_\lambda ^{(1)}({\bf y})}{\partial \rho _\lambda }$ & ${\bf x\neq y}$
\\ 
$\tilde h_{\alpha \lambda }^{(2)}({\bf x,y})=-1$ & ${\bf x=y}$%
\end{tabular}
\right.  \label{hstrip}
\end{equation}

\noindent It must be noticed that, as the stripping is done through a
separation of variables the irreducible two bodies component is not
affected, only the long range behaviour (due to the fixed $N$ constraint) is
isolated. As a consequence, we want to emphasize that $\tilde h_{\alpha
\lambda }^{(2)}$ is the correlation function with a truly irreducible
two-bodies behaviour, not $h_{\alpha \lambda }^{(2)}$. Also, the excluded
volume effects are not altered and the exchange symmetry imposes that $%
\Gamma _{\alpha \lambda }=\Gamma _{\lambda \alpha }$.Therefore, we can
define an $N$- modified full truncated correlation function

\begin{equation}
\tilde t_{\alpha \lambda }^{(2)}({\bf x,y})=\left\{ 
\begin{tabular}{ll}
$\hat t_{\alpha \lambda }^{(2)}({\bf x,y})-\Gamma _{\alpha \lambda }\frac{%
\partial \ln n_\alpha ^{(1)}({\bf x})}{\partial \rho _\alpha }\frac{\partial
\ln n_\lambda ^{(1)}({\bf y})}{\partial \rho _\lambda }$ & ${\bf x}\neq {\bf %
y}$ \\ 
$\hat t_{\alpha \lambda }^{(2)}({\bf x,y})$ & ${\bf x=y}$%
\end{tabular}
\right.  \label{tmod}
\end{equation}

\noindent and verify that it satisfies

\begin{equation}
\int \tilde t_{\alpha \lambda }^{(2)}({\bf x,y})d{\bf x}d{\bf y}=-\Gamma
_{\alpha \lambda }V^2  \label{kernmod}
\end{equation}

\noindent This shows that $\tilde t_{\alpha \lambda }^{(2)}$ is indeed
invertible. Also, $\Gamma _{\alpha \lambda }$ can be written as a functional
of $\tilde h_{\alpha \lambda }^{(2)}$

\begin{equation}
\Gamma _{\alpha \lambda }=-\frac 1{V^2}\int d{\bf x}n_\alpha ^{(1)}({\bf x}%
)\left[ \delta _{\alpha \lambda }+\int d{\bf y}n_\lambda ^{(1)}({\bf y})%
\tilde h_{\alpha \lambda }^{(2)}({\bf x,y})\right]  \label{avcomp}
\end{equation}

\noindent i.e., a sort of averaged compressibility. This recovers the
classical Lebowitz-Percus results \cite{lebper} in a form suitable for our
purposes. Replacing eq. (\ref{tmod}) in (\ref{deldis-us}) and using the
chemical potential definition, it can be obtained that 

\begin{equation}
\delta n_\alpha ^{(1)}({\bf x})=\sum_\lambda \int \tilde t_{\alpha \lambda
}^{(2)}({\bf x,y})\delta \tilde \Phi _\lambda ({\bf y})d{\bf y}
\label{deldis-mod}
\end{equation}

\noindent where $\tilde \Phi _\lambda $ is  the modified external field

\begin{equation}
\delta \tilde \Phi _\lambda ({\bf y})=\delta \Phi _\lambda ({\bf y})+\beta
\delta \mu _\lambda  \label{modfield}
\end{equation}

\noindent If we write that, by definition, 

\begin{equation}
\tilde t_{\eta \alpha }^{(2)-1}({\bf z,x})=\frac{\delta \tilde \Phi _\eta (%
{\bf z})}{\delta \tilde n_\alpha ^{(1)}({\bf x})}=-\tilde c_{\eta \alpha
}^{(2)}({\bf z,x})+\frac{\delta _{\alpha \eta }\delta ({\bf z-x})}{\tilde n%
_\alpha ^{(1)}({\bf x})}  \label{invt-mod}
\end{equation}

\noindent the $N$-modified functions $\tilde t_{\eta \alpha }^{(2)-1},\tilde 
t_{\alpha \lambda }^{(2)}$ obey eq. (\ref{ozfunct}) and, therefore, $\tilde h%
_{\alpha \lambda }^{(2)}$ and $\tilde c_{\eta \alpha }^{(2)}$ are linked by
an OZ equation with $\tilde c_{\eta \alpha }^{(2)}$ playing the role of an $N
$-modified direct correlation function.

Now we define an $N$-modified intrinsic free energy functional by

\begin{equation}
\beta {\sl \tilde F}\left[ \{\tilde n_\alpha ^{(1)}\}\right] =\left\langle
\beta (K+U-\ln P_N)\right\rangle  \label{intr-f}
\end{equation}

\noindent where $K,U,P_N$ are the kinetic, potential energy and $N$-bodies
probability distribution function in the CE respectively with the average
being a CE one and minimize the functional

\begin{equation}
\beta {\sl \tilde \Omega }\left[ \{\tilde n_\alpha ^{(1)}\}\right]
=-\sum_\alpha \int \tilde n_\alpha ^{(1)}({\bf x})\tilde \Phi _\alpha ({\bf x%
})d{\bf x+}\beta {\sl \tilde F}\left[ \{\tilde n_\alpha ^{(1)}\}\right]
\label{tpot-mod}
\end{equation}

\noindent It is easy to prove, using the Gibbs-Bogoliubov inequality \cite
{revliq}, that the minimum is reached when the equilibrium profiles $\{%
\tilde n_\alpha ^{(1)}\}$ are used. This completes the proof of the
extension of the HKSM theorem to the CE. It can also be seen that the $N$%
-modified direct correlation function hierarchy $\{\tilde c^(s)\}$ can be
started by

\[
\tilde c_\alpha ^{(1)}({\bf x})=\beta \frac{\delta {\sl \tilde F}%
^{exc}\left[ \{\tilde n_\alpha ^{(1)}\}\right] }{\delta \tilde n_\alpha
^{(1)}({\bf x})}
\]

\noindent And in particular, we have

\[
\tilde c_{\alpha \lambda }^{(2)}({\bf x,y})=\frac{\delta \tilde c_\alpha
^{(1)}({\bf x})}{\delta \tilde n_\lambda ^{(1)}({\bf y})}=\frac{\delta
_{\alpha \lambda }\delta ({\bf x-y})}{\tilde n_\alpha ^{(1)}({\bf x})}-\frac{%
\delta \tilde \Phi _\alpha ({\bf x})}{\delta \tilde n_\lambda ^{(1)}({\bf y})%
}
\]

\noindent Therefore, we have proved that all of our $N$-modified functions
can be generated by functional differentiation with the formal structure of
eqs. (\ref{dismod-us}), (\ref{corrmod-us}) and (\ref{dcorr-us}).

Summarizing, we have proved that the HKSM\ theorem of the GCE\ can be
extended to the CE and, in the process of doing that, we have defined a new
set of $N$-modified distribution and correlation functions that can be built
using the same rules and formal structure of the conventional canonical
functions with the proviso that the set of modified external fields must be
used. A paper with a more detailed derivation together with some
applications is in preparation.

We acknowledge support from the National Science Foundation through grants
CHE-95-13558, Epscor OSR-94-52893, by the DOE-EPSCoR grant DE-FCO2-91ER75674
and CONICET grant PIP 0859/98.

\bibliographystyle{plain}
\bibliography{}

\end{multicols}

\end{document}